\documentclass[12pt,epsf]{article}
\newlength{\abstractwidth}
\usepackage[dvips]{graphicx}
\usepackage{subfigure}

\catcode`\@=11
\@addtoreset{equation}{section}
\def\theequation{\arabic{section}.\arabic{equation}}
\catcode`@=12
\relax
\renewcommand{\thefootnote}{\fnsymbol{footnote}}
\renewcommand{\thanks}[1]{\footnote{#1}} 
\newcommand{\starttext}{
\setcounter{footnote}{0}
\renewcommand{\thefootnote}{\arabic{footnote}}}
\renewcommand{\theequation}{\thesection.\arabic{equation}}
\newcommand{\be}{\begin{equation}}
\newcommand{\bea}{\begin{eqnarray}}
\newcommand{\eea}{\end{eqnarray}}
\newcommand{\beq}{\begin{equation}}
\newcommand{\ee}{\end{equation}}
\newcommand{\eeq}{\end{equation}}

\def\<{\langle}

\def\ba{\begin{eqnarray}}
\def\ea{\end{eqnarray}}

\begin{document}
\renewcommand{\theequation}{\thesection.\arabic{equation}}
\begin{titlepage}
\bigskip
\rightline{SU-ITP-08-12} 
\rightline{OIQP-08-06}
\rightline{UCB-PTH-08-60}

\bigskip\bigskip\bigskip\bigskip

\centerline{\Large \bf {Future Foam}}

\bigskip\bigskip
\bigskip\bigskip

\centerline{\it Raphael Bousso\,${}^{1}$, Ben Freivogel\,$^{1}$, 
Yasuhiro Sekino\,$^{2,3}$, Stephen Shenker\,$^{3}$, }
\centerline{\it Leonard Susskind\,$^{3}$, I-Sheng Yang\,$^{1}$, 
and Chen-Pin Yeh\,$^{3}$} 
\bigskip

\bigskip
\centerline{${}^1\;$ Department of Physics and Center for Theoretical
 Physics, University of California, Berkeley,}
\centerline{and  Lawrence Berkeley National Laboratory,
Berkeley, CA 94720, U.S.A.}
\centerline{${}^2\;$ Okayama Institute for Quantum Physics, 
Okayama 700-0015, Japan}
\centerline{${}^3\;$ Physics Department, Stanford University, 
Stanford, CA  94305, U.S.A.}

\bigskip

\bigskip

\bigskip

\bigskip

\begin{abstract}
We study pocket universes which have zero cosmological constant and
non-trivial boundary topology.  These arise from bubble collisions 
in eternal inflation.  Using a simplified dust model of collisions 
we find that boundaries of any genus can occur.
Using a radiation shell model we perform analytic studies in the thin 
wall limit to show the existence of geometries with a single toroidal 
boundary.  We give plausibility arguments that higher genus boundaries 
can also occur.  In geometries with one boundary
of any genus a timelike observer can see the entire boundary.  
Geometries with multiple disconnected boundaries can also occur.  
In the spherical case with two boundaries the boundaries are separated 
by a horizon. Our results suggest
that the holographic dual description for eternal inflation, proposed by
Freivogel, Sekino, Susskind and Yeh,
should include summation over the genus of the base space of the dual
conformal field theory.  We point out peculiarities
of this genus expansion compared to the string perturbation series.
\end{abstract}
\end{titlepage}
\starttext \baselineskip=18pt \setcounter{footnote}{0}

\section{Introduction}

The nonperturbative definition of
string theory in asymptotically flat or anti-de Sitter (AdS)
spacetimes is now basically understood. 
Matrix theory~\cite{BFSS} and AdS/CFT correspondence~\cite{adscft} 
provide concrete non-perturbative (holographic)
formulations of quantum gravity in terms of non-gravitational gauge theories.
These theories taught us many things. For instance, the fact that
formation and evaporation of a black hole is mapped
to a manifestly unitary process in gauge theory makes us strongly
believe that information is not lost in black holes. 

On the other hand, it is not yet clear how to define an exact quantum
theory for a cosmological, or inflating, spacetime. The main source of
difficulty seems to be the fact that there is no obvious asymptotic
region where interactions are turned off. 

Finding a non-perturbative framework for cosmology is
especially important because of the existence of
the string landscape~\cite{landscape}: string theory
contains a large number of vacua including metastable de Sitter vacua.
Metastability is an approximate concept. Even though
there is strong evidence for the existence of the landscape, which
is obtained from the low-energy effective theory analysis, the meaning
of these metastable vacua is not totally clear until
we have an exact theory.

In the landscape, bubbles (universes in different vacua, or 
``pocket universes'') are created by tunneling. Eternal
inflation generically occurs. Here the false vacuum inflates 
so fast that bubbles of true vacuum cannot percolate and 
the physical volume of the space remains dominated by the 
false vacuum forever. Infinitely many bubbles 
are produced, and the volume inside each bubble is also 
infinite. It is not known how to regulate these infinities. 
This is related to the measure problem. 
If we can find a non-perturbative framework, it may give some
clues about this problem.

In~\cite{FSSY}, a holographic dual description for eternal 
inflation was proposed. The authors consider an FRW
universe with zero cosmological constant (c.c.) created
by tunneling from de Sitter space. The tunneling is described by
the Coleman-De Luccia (CDL) instanton~\cite{CDL}, which tells
us that the FRW universe is an open universe whose constant time 
slices are 3-dimensional hyperboloids. The dual theory is
a conformal field theory (CFT) defined on $S^2$ at the boundary 
(spatial infinity) of the 3-hyperboloid~(Figure~1).
\begin{figure}
\center
\includegraphics[scale=.33]{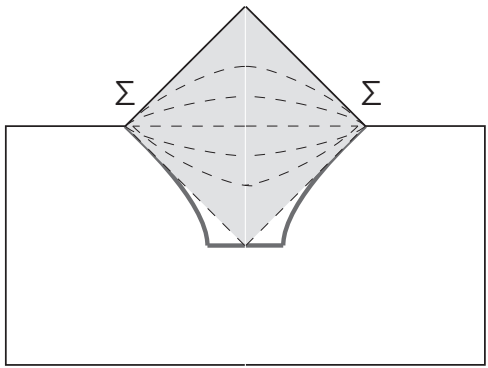}
\caption{Bubble nucleation in de Sitter space. Thick lines
are the domain wall between true and false vacuum.  
Shaded region is an open FRW universe, in which  
constant time slices (dotted lines) are $H^3$.
Vacuum energy of the true vacuum is assumed to be zero 
asymptotically, FRW universe has future asymptotics of
flat space (the ``hat'').
The dual CFT is defined on the boundary $\Sigma$ of $H^3$, 
which is $S^2$. (Note that this is the ``doubled'' Penrose diagram; 
the two points denoted by $\Sigma$ are on the same $S^2$.)}
\label{fig-CDL}
\end{figure}

In this ``FRW/CFT duality,'' the dual theory  
contains 2-dimensional gravity (the Liouville field). 
One may wonder why gravity is not decoupled on the boundary, as 
in AdS space where it is fixed with a boundary condition. 
The reason for non-decoupling
is that the FRW universe is embedded in de 
Sitter space. Our boundary is alternately regarded as the bubble 
wall at future infinity of de Sitter space (see Figure~1). 
In de Sitter space, fluctuations produced at early time 
are stretched by inflation and cannot be smoothed out by 
late-time fluctuations, so the fluctuations at two points
remain correlated after those points 
go out of causal contact (see e.g.~\cite{linde}). 
The gravity fluctuations on the boundary of the FRW universe is 
of the same origin. Indeed, the graviton correlator 
computed in~\cite{FSSY} using the Euclidean prescription remains 
finite as the points approach the boundary of the FRW universe. 
A boundary such as this one where gravity is
not decoupled is called a ``warm'' boundary~\cite{censustaker}, 
as opposed to the ``cold'' boundary of (global) AdS space. 
In most proposals for a holographic duality for inflation, such as 
the dS/CFT correspondence~\cite{dscft} and the 
dS/dS correspondence~\cite{dsds}, 
gravity is not decoupled.


In addition to the above perturbative argument, 
bubble collisions which are inevitable 
in eternal inflation indicate
that the boundary geometry is fluctuating.
Consider a collision of two bubbles of the same 
vacuum, for which no domain wall remains after the collision.
The bulk space will approach a smooth geometry. If the 
c.c. of the vacuum is zero, 
a timelike observer can see the whole space inside the bubbles 
eventually; the geometry contains only one ``hat'' (future 
asymptotics of flat space).
However, the boundary geometry will be deformed from a 
perfect sphere.
It is conjectured that a bubble collision corresponds to
an instanton in the dual theory~\cite{FSSY}.

In this paper, we point out that the boundary is not only
``metrically warm,'' but is also ``topologically warm.''
A universe with a non-trivial boundary topology can arise from 
bubble collisions. We can easily imagine a ``ring'' that appears
as a result of collisions of three or more bubbles (Figure~2). 
The region ``in the middle'' does not close up if its size
is larger than the horizon scale. Even though
the wall of the true vacuum moves outwards, the inflation
of de Sitter space surpasses it. 
A timelike observer in the true vacuum (the ``bulk'' of the torus) can 
eventually see the whole boundary, as we discuss in section 2.
Boundaries with general higher topologies will also be present. 
\begin{figure}
\center
\includegraphics[scale=.25]{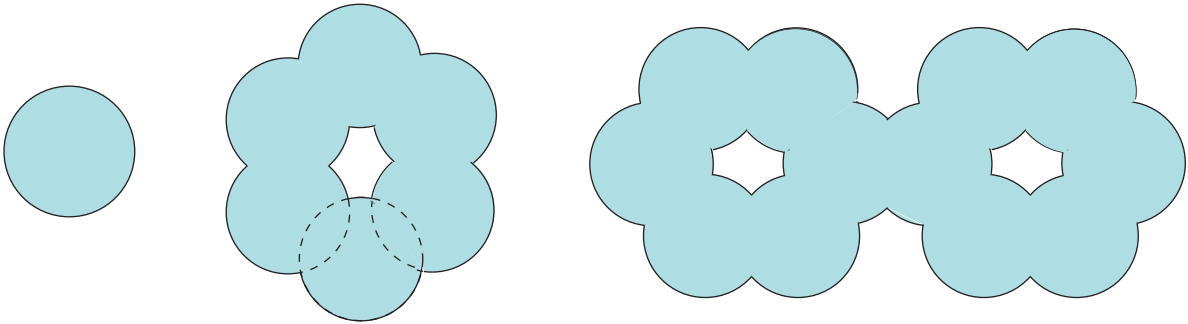}
\caption{Boundary of the true vacuum region with genus 0, 1, 2.
In the middle figure, we have indicated by dotted lines
where domain walls would be if there were no collisions.}
\label{fig-ring}
\end{figure}

Non-trivial topologies are suppressed by powers of the 
nucleation rate $\gamma$, compared to the spherical topology, 
and may not be important observationally. 
However to define the holographic theory we should include
everything that a single observer can see. This means that
we should sum over topologies of the 2-dimensional space
on which the CFT is defined. 

Summation over topologies is reminiscent of string perturbation theory,
which is an asymptotic expansion. The string perturbation series does 
not converge, and important objects, such as D-branes, cannot be seen 
in a perturbative expansion. 
We might wonder whether a similar thing happens in the genus expansion in our 
dual theory. 
We will estimate the growth of terms in the series by taking the
sum over bubbles, and argue that
here the series converges.

We also note that bubble collisions can produce spacetimes
with multiple boundaries. This happens when the bubbles form a 
``shell,'' for example. If two boundaries are accessible to
a single timelike observer, it would be confusing 
in terms of the dual theory. For the case of two spherical
boundaries, we can show that the two boundaries are causally
disconnected. It would be very interesting to know whether this
 generalizes to cases with more complicated topology, so that a given observer can only see a single connected boundary.

This paper is organized as follows: In section 2, 
we explain that universe with non-trivial topology can 
be produced by bubble collisions. We first give a general argument
based on the assumption that bubble walls turn into a wall of dust
after colliding. We then perform analytic
study in the thin-wall approximation,
for the two limiting cases with torus topology. 
We first consider bubbles aligned on a straight line 
with equal spacing, and obtain asymptotic 
geometry after collisions. We then consider a ``coarse grained'' 
version, in which we approximate the domain wall between
the true and the false vacuum regions by a smooth torus.   
In section 3, we show that there can be a universe with multiple 
boundaries. We study the case where the true vacuum is inside
a shell-like region, and see that a singularity develops
inside the shell, leaving two boundaries causally disconnected. 
In section 4, we discuss implications of the non-trivial
boundary topology for the dual theory. We study 
large order behavior of the genus expansion, and discuss
possible interpretation in the dual CFT.

\section{Boundary with non-trivial topology}

We shall consider the simplest setting in this paper: Gravity is coupled
with a scalar field whose potential has two minima, 
a false vacuum with positive c.c.  and a true vacuum with zero c.c.
We consider four spacetime dimensions. 

If the space is filled with false vacuum, 
bubbles of true vacuum will be nucleated with a rate $\gamma$
which is calculated from the CDL instanton~\cite{CDL}.
We are interested in diagnosing the topology of the boundary between the true vacuum and the false vacuum at conformal infinity. Before taking true vacuum bubbles into account, conformal infinity of de Sitter space is a 3-sphere. When a bubble of true vacuum nucleates, it eats up a ball out of the conformal infinity of de Sitter space. The size of the ball depends on the time of nucleation. The boundary of the true vacuum region is the boundary of the ball, a 2-sphere.  When two bubbles of true vacuum collide in de Sitter space the region of de Sitter conformal infinity which is removed is simply given by superposing the two balls from each nucleation. The geometry inside the balls may be complicated, and depends on the physics of what happens when the walls collide. However, the boundary between de Sitter space and the true vacuum is simply given by superposing balls of different sizes, one for each true vacuum bubble, and then looking at the boundary of this region.
It is possible that the nucleated bubbles collide and 
form a ring (or ``chain of pearls''; see Figure~\ref{fig-ring}) where
the true vacuum is inside a torus, or in a similar fashion anything 
with higher genus.

An important question is whether a single observer inside the true vacuum can see the entire boundary. To answer this question, it is necessary to construct the geometry inside the true vacuum regions. As we mentioned above, the geometry depends on the physics of what happens when the domain walls collide.
We first demonstrate the existence of a boundary
of arbitrary genus, all of which is visible to a single observer in the true vacuum region,
 using a simplified ``dust'' model in section 2.1.
  Then
using a somewhat more 
realistic ``radiation shell'' model we construct the smooth
solution in the simplest case, a torus,
in Section 2.2 and 2.3. Finally in Section 2.4, we make a 
conjecture about the smooth geometry inside a boundary of any genus 
based on our torus solution.

\subsection{Chain of pearls with dust walls}

For the moment, we assume that when two bubbles collide, their domain walls annihilate into
a  (2+1) dimensional wall of dust. In other words, we assume that there is a type of ``domain wall" with equation of state $P=0$ separating the two regions of true vacuum after the bubbles collide.
The resulting geometry for a single collision
was constructed in ~\cite{BFY}.  The entire true vacuum region is within the backward lightcone of a single observer.  With more collisions, as 
long as the resulting dust walls do not cross each other, we can 
iterate the same solution and show that the entire interior region 
is causally connected.  The non-trivial question is, can the dust 
walls stay away from each other for the collisions necessary to make
true vacuum regions of various topologies? We will find that connected boundaries
of arbitrary topology can be constructed using this simple procedure, but disconnected boundaries cannot be constructed in this way.
 
Consider a given true vacuum bubble which collides with several other true vacuum bubbles. We need to know whether the dust walls from these various collisions intersect each  other. In the thin wall approximation, the interior of each bubble is the Milne universe out to the dust walls, with metric
\be
\label{milne}
ds^2 = - dt^2 + t^2 ds^2_{H^3}
\ee
where $ds^2_{H^3}$ is the metric on 3-dimensional hyperbolic space. The conformal boundary of $H^3$ is a 2-sphere. This is the boundary between de Sitter space and Minkowski space for a single bubble.

Now consider collisions. A given collision destroys part of the original $S_2$ boundary. At conformal infinity, when two true vacuum balls overlap, the part of each $S_2$ which is inside the other ball is destroyed. Focusing on a given bubble, collisions punch holes in the boundary $S_2$. These holes are the interiors of circles, because the overlapping $S_2$'s intersect in a circle. The destroyed pieces of domain wall have been converted to dust walls. Our concern is whether these dust walls intersect.

Studying the dynamics of the dust walls gives the result that at late time in the coordinates (\ref{milne}) the dust wall asymptotes to the minimal surface inside $H^3$ whose boundary is the intersection circle  \cite{BFY} . This minimal surface is simply an $H^2$ with unit radius. (At earlier times, the dust wall is not yet a minimal surface, but we are interested in late time because  the dust wall extends maximally far into the space at $t \to \infty$.) Now if a given bubble collides with several other bubbles, then there are several dust walls emanating from the intersection circles. The minimal surfaces (dust walls) intersect if and only if the different intersection circles intersect each other.

So we have a simple rule for building a large class of solutions in which the entire true vacuum region is causally connected. Start with de Sitter conformal infinity, which is an $S_3$, and put down balls of true vacuum of any size in any location. The boundaries of the balls are $S_2$'s, which when they intersect generically intersect in circles. {\it The only rule is that the intersection circles cannot intersect each other.} 

(It is not obvious from our analysis here, but the dust walls intersect if and only if the collision $H_2$'s intersect. When this happens black holes generically form \cite{moss}, so the analysis becomes difficult no matter what assumption we make for the physics of the collision.)

Now what kinds of interesting boundary topologies can be constructed in this way?
It is possible to connect geometries with arbitrary {\it connected} boundary using this construction.
For example, 
we can construct a true vacuum region of torus topology, as shown in 
 Figure \ref{fig-ring}. 
Since the true vacuum regions overlap only pairwise, the intersection circles are all well separated from each other, and our construction works.

On the other hand, it is not possible to construct geometries with disconnected boundaries using this technique. The boundary of the true vacuum region is constructed out of a number of $S_2$'s  which
are glued together along circles. The interiors of the circles are holes in the $S_2$'s.  Since the circles do not intersect in this construction, it is possible to get from one point along a given holey $S_2$ to any other point.  Also, it is possible to move from one holey $S_2$ to one connected to it by moving across the gluing circle. Therefore, the boundary of the true vacuum region is connected.


\subsection{Sequence of collisions}

We now turn to a more detailed analysis of bubble collisions where we make
the somewhat more realistic assumption
that all energy is liberated in a shell of radiation.

For simplicity, we would like to start with as many symmetries as 
possible.  A de~Sitter space with one bubble has SO(3,1) symmetry;
this is inherited from the spherical symmetry SO(4) 
of the Euclidean CDL instanton~\cite{CDL}. 
When there are two bubbles, the direction connecting the
centers of the bubbles singles out a preferred direction, but
there is SO(2,1) residual symmetry.  
When there are four or more bubbles, there is generically
no residual symmetry. However, if nucleation sites are on
the same spacelike geodesic ( the great circle of the minimal 
$S^3$ ), SO(2,1) symmetry is preserved.  In addition, the circle
of nucleation points gives us a discrete subgroup of $U(1)$, so 
we have $SO(2,1)\times U(1)$ which contains a torus.

We will use coordinates with manifest SO(2,1) 
symmetry~\cite{moss2, moss}. De Sitter space can be written as 
\begin{equation}
 ds^2= - f(t)^{-1}dt^2+f(t)dz^2 +t^2 dH_2^2,
\label{h2metric}
\end{equation}
with 
\begin{equation}
 f(t)=1+{t^{2}\over l^2}
\label{fdS}
\end{equation}
where $l$ is the de Sitter radius 
(the horizon size)\footnote{The de Sitter metric (\ref{fdS}) 
is obtained by parameterizing the
embedding coordinates in R$^{4,1}$ (which satisfy
$-X_0^2+\sum_{a=1}^{4}X_a^2=l^2$)
as follows: $X_{a}=t\, H_a$ 
(where $a=0,1,2$, and $-H_0^2+H_1^2+H_2^2=-1$), 
$X_3=\sqrt{l^2+t^2}\, \cos (z/l)$, 
$X_4=\sqrt{l^2+t^2}\, \sin (z/l)$. The configuration considered
here is invariant under SO(2,1) acting on the $X_0, X_1, X_2$ space. 
The metric (\ref{fdS}) does not cover the whole de Sitter. To
study the trajectory in the directions other than $z$, it is more 
convenient to use global coordinates (\ref{dSglobal}).}, 
and $0\le z\le 2\pi l$. 
The bubbles are nucleated at time $t=0$, along the circle in the
$z$ direction.  
For simplicity, we assume that the nucleation
sites are evenly spaced with distance $2\Delta z$. (We 
are assuming $N=\pi l/\Delta z$
is an integer.)  
The initial size $r_0$ of a bubble is determined by 
the parameters of the scalar potential.
\begin{figure}
\center
\includegraphics[scale=.33]{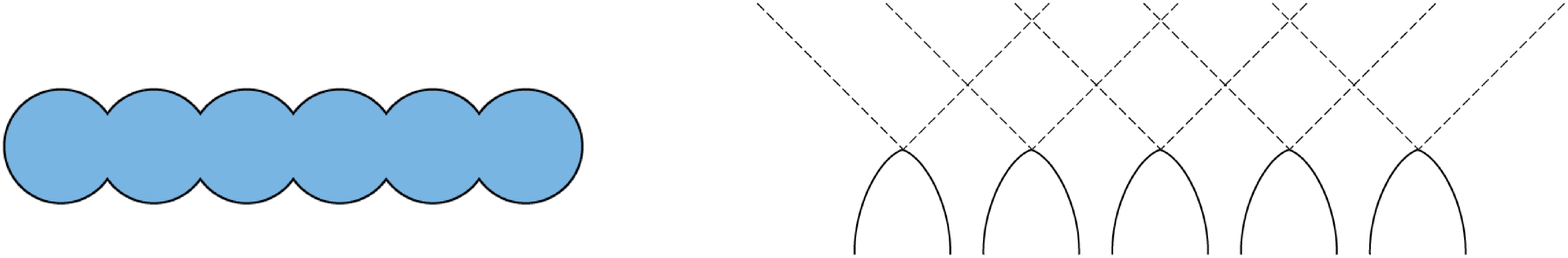}
\caption{Left: True vacuum (shaded region) inside a torus 
(The left and the right ends are identified). This configuration is 
produced by bubble collisions. Right: Trajectory of domain walls 
in the $(t,z)$ plane. (Horizontal ($z$) direction is periodically
identified.) Bubble walls (solid lines) collide and emit radiation 
(dotted lines). A  shell of radiation collides with other shells
an infinite number of times.}
\label{fig-chain}
\end{figure}

The profile of bubble walls in the ($t,z$) plane will be
as depicted in Figure~\ref{fig-chain}. 
In the thin-wall limit, geometry in each bubble is flat.
We parametrize flat space in such a way that 
$H^2$ factor is manifest, which is of the form (\ref{h2metric}) 
with $f=1$. We patch it to de Sitter space on the domain wall. 
Since the metric component along the wall should
be continuous across the wall, 
the coordinate $t$ (which sets the scale for the $H^2$ metric)
should have the same value when we approach 
the domain wall from either side. The coordinate $z$ for the flat 
and the de Sitter side will be different. 
The trajectory of the domain wall 
is determined by the Israel junction condition
once the equation of state for the domain wall is given. 

\begin{figure}
\center
\includegraphics[scale=.33]{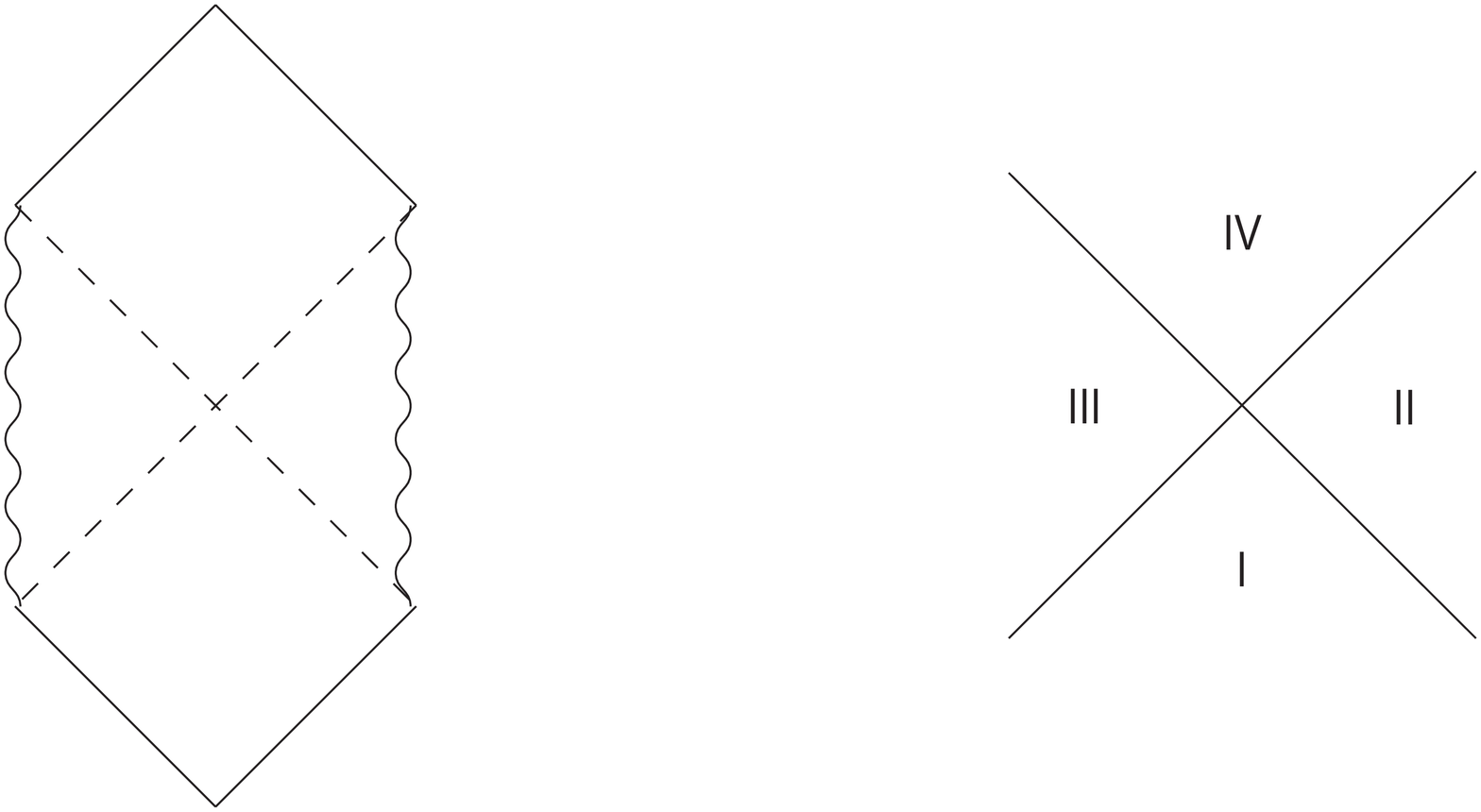}
\caption{Left: Causal structure of the hyperbolic Schwarzschild
geometry, whose metric is given by (\ref{h2metric}) with
 (\ref{fzerocc}). The ($t,z$) plane is shown; on each point, 
$H^2$ is attached. There are timelike singularities at $t=0$. 
The dotted lines are null planes at $t=t_1$. We will only use 
the $t>t_1$ region, which is the upper diamond. Right: Collision 
of null domain walls. The $(t,z)$ plane is
divided into four regions.}
\label{fig-regions}
\end{figure}

Bubble collisions occur along an $H^2$. To find the metric after  
collision, we make an assumption following~\cite{moss2, moss}: 
When two bubbles 
collide, the bubble walls disappear instantaneously and turn into radiation. 
Radiation follows a light-like trajectory in the $(t,z)$ space. 
The geometry in the region behind the radiation differs from
flat space in general, since the radiation carries away some
energy. The metric will be of the form (\ref{h2metric}) with 
\begin{equation}
f(t)=1-{t_{1}\over t}.
\label{fzerocc}
\end{equation}
This is the most general zero-c.c. geometry with $H^2$ symmetry;
it is the hyperbolic version of the Schwarzschild geometry.
The causal structure of this metric 
is given in Figure~\ref{fig-regions}. 
There is a 
time-like singularity at $t=0$, but as we
will see, only the $t>t_{1}$ ($>0$) region is relevant for us.

The parameter $t_{1}>0$ is determined by the following
condition \cite{moss}.
To make the formula simple, let us ignore the initial size of each bubble
so that the bubble walls are moving at the speed of light. 
When two light-like domain walls collide and emit two 
light-like domain walls (i.e. walls of radiation),
spacetime is divided into four regions. We will label them
I, II, III, IV in the way depicted in Figure~\ref{fig-regions}, 
and denote the function $f(t)$ in those regions by $f_I(t)$ and so on. 
At the time of the collision $t=t_{*}$ (which is common in all 
the regions), they satisfy
\begin{equation}
f_{I}f_{IV}=f_{II}f_{III}.
\label{conservation}
\end{equation}
This is essentially the energy conservation condition~\cite{maeda}. 
Substituting de Sitter metric 
(\ref{fdS}) for $f_{I}$ and flat metric $f=1$ for $f_{II}$ and $f_{III}$, 
we find 
\begin{equation}
 f_{IV}=1-{t_{1}\over t_*}= \left(1+ {t_{*}^2\over l^2}\right)^{-1}.
\label{f4}
\end{equation}
The region IV is the $t\ge t_{*}$ part of the metric (\ref{fzerocc}), 
and the above equation implies $t_{*}>t_{1}$. Thus, there is no singularity
in region IV. 

The collision time $t_{*}$ is determined by the initial condition.
In terms of the conformal coordinate $T$,  
defined by $T=\int dt/(lf_I(t))=\arctan(t/l)$, 
bubbles separated by a distance
$2\Delta z$ collide in a time $\Delta T=\Delta z/l$. 
If the separation at the nucleation time $t=0$ is small,
$\Delta z\ll l$, we can approximate $f_I(t)\sim 1$,
and get $t_{*}\sim l\Delta T=\Delta z$.  
In this limit, the condition (\ref{f4}) becomes 
\begin{equation}
t_{1}={t_{*}^3\over l^2+t_{*}^2}
\sim  {t_{*}^3\over l^2}\sim {(\Delta z)^3\over l^2}.
\end{equation}

Geometry of the region IV is maximally curved at the earliest
time, $t=t_{*}$, where the deviation of $f_{IV}(t_*)$ from 1
is 
\begin{equation}
 {t_{1}\over t_*}={(\Delta z)^2\over l^2}.
\end{equation}
This can be made arbitrarily small by making $\Delta z/l$ small. 

As we see from Figure~\ref{fig-chain}, a wall of radiation 
collides with another one which comes from the neighboring collision. 
Again, walls of radiation are emitted at the collision, and 
the metric behind the radiation is changed. 
This process will be repeated infinite times. 

The geometry after $n$ such collisions is obtained by 
using the junction condition iteratively. 
Let us call $f_{n}$ the function $f$ after $n$-th collision
(which means this $f_{II}=f_{III}=f_{0}$ and 
$f_{IV}=f_{1}$) and write it 
\begin{equation}
 f_{n}=1-{t_{n}\over t}.
 \label{eq-sch}
\end{equation}
Also we define $t_{*(n)}$ to be the time of $n$-th collision
(the $t_*$ above is $t_{*(1)}$).  
The condition (\ref{conservation}) gives us
\begin{equation}
\left(1+\frac{t_n}{t_{*(n+2)}}\right)
\left(1+\frac{t_{n+2}}{t_{*(n+2)}}\right)=
\left(1+\frac{t_{n+1}}{t_{*(n+2)}}\right)^2~. 
\label{recursion1}
\end{equation}
There is another condition which says that the coordinate
distance traveled by the light between $n$-th and $n+1$-th
collisions is equal to the one between $n+1$-th and $n+2$-th
collisions: 
\begin{equation}
t_{*(n+2)}-t_{*(n+1)}+t_n\ln
\left(\frac{t_{*(n+2)}-t_n}{t_{*(n+1)}-t_n}\right)
=t_{*(n+1)}-t_{*(n)}+t_n\ln\left(
\frac{t_{*(n+1)}-t_n}{t_{*(n)}-t_n}\right).
\label{recursion2}
\end{equation}

These recursion relations simplify when the metric is close to
flat. If $t_{n}/t_{*(n)}$, $t_{n}/t_{*(n+1)}$, 
$t_{n}/t_{*(n+2)}$ are all much smaller than 1, the leading
part of (\ref{recursion1}) and (\ref{recursion2}) becomes
\begin{eqnarray}
t_{n+2}-t_{n+1}&=&t_{n+1}-t_{n}, \\ 
t_{*(n+2)}-t_{*(n+1)}&=&t_{*(n+1)}-t_{*(n)}.
\end{eqnarray}
Together with our initial conditions, $t_{0}=t_{*(0)}=0$, 
$t_{1}=(\Delta z)^3/l^2$, and $t_{*(1)}=\Delta z$,
these give us
\begin{equation}
 t_{n}=n {(\Delta z)^3\over l^2}, \quad t_{*(n)}=n \Delta z.
\end{equation}
When $\Delta z \ll l$, the deviation from the flat metric is always 
small, $t_{n}/t_{*(n)}= (\Delta z /l)^2 \ll 1$, so our approximation 
is consistent.  

\begin{figure}
   \centering
   \includegraphics[width=2.5in]{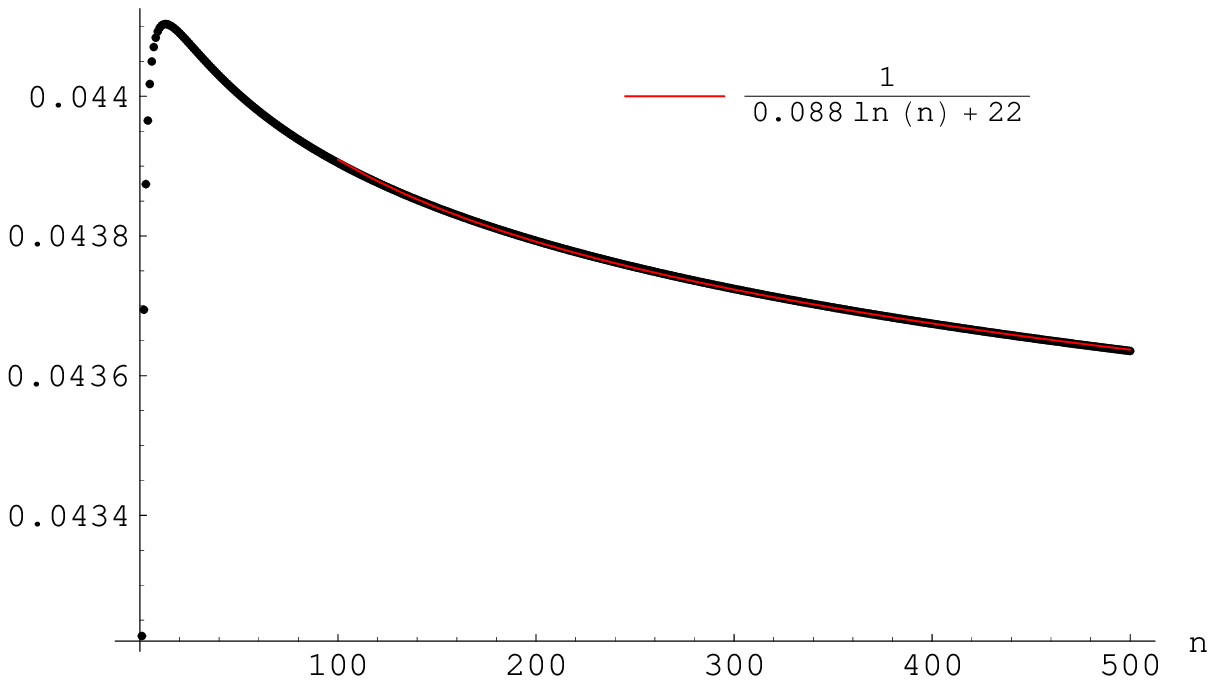}
\includegraphics[width=2.5in]{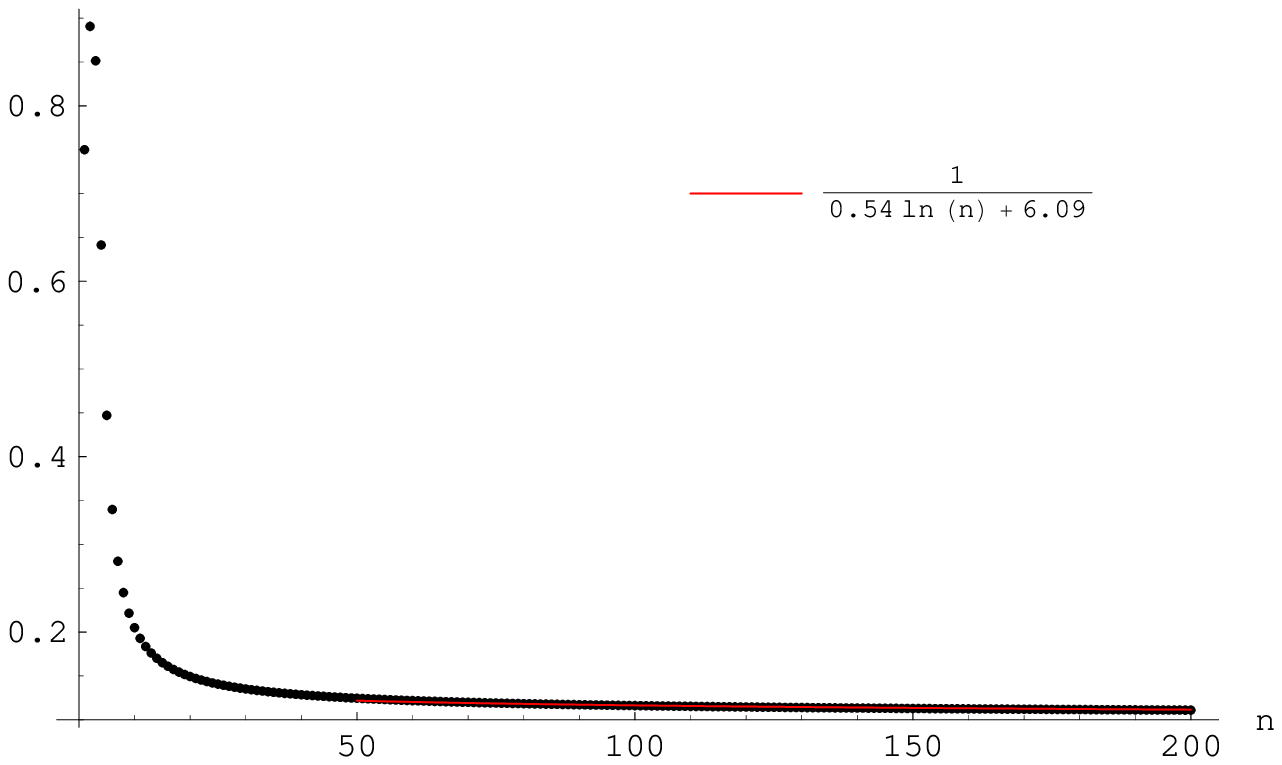}
   \caption{The function $t_n/t_{*(n)}$ obtained by iteration,
for the collision of $N$ bubbles. Left: $t_n/t_{*(n)}$ for $N=15$;
Right: $t_n/t_{*(n)}$ for $N=3$. The solutions are fitted by 
$\frac1{a\ln(n) +b}$ in the $n>100$ and $n>50$ regions, 
respectively.}
\label{fig-iteration}
\end{figure}

We can compute corrections by substituting this leading order solution
into (\ref{recursion1}), (\ref{recursion2}) and solving
them perturbatively in $\Delta z/l$. 
The function $t_{n}/t_{*(n)}$ decreases at subleading 
order\footnote{This expression is valid when $\ln n$ is sufficiently
small. At some point the error accumulates and this lowest order
approximation breaks down.} as 
$t_{n}/t_{*(n)}\sim (\Delta z /l)^2 (1-2(\Delta z /l)^4\ln n)$.
The geometry asymptotically approaches flat space locally, 
although the logarithmic rate of approach is slower than 
the rate $1/t$ for the ordinary collision of two bubbles.
Numerical solutions obtained by iteration  
(without assuming $f_n\sim 1$) indeed shows that the function 
$t_n/t_{*(n)}$ decreases logarithmically for 
large $n$ (See Figure~\ref{fig-iteration}).  We can also see
that the maximum value of $t_n/t_{*(n)}$ decrease as we
make $\Delta z/l=\pi/N$ small, so our perturbative analysis can 
always work by making this control parameter small. 

There is an useful physical picture for the above analysis.  
If there is no incoming radiation, Eq.~(\ref{eq-sch}) will approach 
flat space locally.  The leading order analysis shows that for an observer, 
the radiation walls keep arriving at equal time intervals, therefore 
they serve as a constant reminder of non-flatness.  The perturbative
correction and the simulation suggest that radiation walls actually 
arrive at {\it increasing} time intervals, which guarantees flatness
at future infinity. 

Taking the smooth ($\Delta z/l\ll1$) limit, we have exact $SO(2,1)
\times U(1)$ symmetry, the boundary of the flat region has two 
non-contractible circles.  In the global coordinates for de Sitter,
\begin{equation}
 ds^2=-d\hat{t}^2 +\cosh^2\hat{t}\left(d\alpha^2+
\cos^2\alpha dz^2 +\sin^2\alpha d\theta_2^2\right),
\label{dSglobal}
\end{equation} 
the above solution looks like the future lightcone of the $z$ axis 
circle at $\hat{t}=0,\alpha=0$.  In general we can also consider a 
circle at $\hat{t}=\hat{t}_0>0$.  At future infinity,
the bubble wall (radial light ray from $\alpha=0$) 
reaches $\alpha=\alpha_1=\arcsin(1/\cosh \hat{t}_0)$.
One of the circles of the torus has radius
$r_2=\cosh\hat{t}\sin \alpha_1$. This is the circle
($\theta_2$ direction) contained in one bubble.
The other circle of the torus is in the $z$-direction which 
traverses many bubbles, and has radius $r_1=\cosh\hat{t}\cos\alpha_1$, 
Asymptotic ratio of the two radii is
\begin{equation}
 {r_2\over r_1}=\tan \alpha_1={1\over \sinh\hat{t}_0}.
\end{equation}
We established the asymptotic local flatness for the $\hat{t}=0$ 
case, in which even though $r_2$ becomes infinite but $r_1$ remains 
finite and equals to the horizon size $r_1=l$.  For $\hat{t}_0>0$, 
$r_1$ also grows to infinity. Each source of radiation (the collision
$H_2$) is moving away from each other. This should make 
the approach to flat space faster. 
We should be able to patch flat space to de Sitter across a toroidal 
domain wall. We will study this in the next subsection.

\subsection{Coarse grained smooth torus}

We will construct the smooth torus solution suggested in the previous 
subsection, which is more general since it has only $U(1)\times U(1)$ 
symmetry. 

To make the symmetry manifest, we express the global de Sitter space as
\begin{equation}
ds^2=\frac{(r_2^2-l^2)dr_1^2+(r_1^2-l^2)dr_2^2-2r_1r_2dr_1dr_2}
          {r_1^2+r_2^2-l^2}
          +r_1^2d\theta_1^2+r_2^2d\theta_2^2~,
\label{eq-torusDS}
\end{equation}
where $r_1^2+r_2^2\ge l^2$, and $0\le \theta_1, \theta_2\le 2\pi$.
\footnote{This metric is obtained by parameterizing the embedding 
coordinates in $R^{4,1}$ as
$X_0=\pm \sqrt{r_1^2+r_2^2-l^2}$, $X_1=r_1\cos\theta_1$, 
$X_2=r_1\sin\theta_1$, $X_3=r_2\cos\theta_2$, $X_4=r_2\sin\theta_2$.
The relation of these coordinates to the usual global time, defined by 
$ds^2=-d\hat{t}^2+\cosh^2\hat{t} d\Omega_3^2$, is: 
$\sinh\hat{t} =\sqrt{r_1^2+r_2^2-l^2}/l$.}

For the interior flat space, we start from the Minkowski space with 
a manifest $H_1$,
\begin{equation}
ds^2=-dt^2+t^2d\xi^2+dr_2^2+r_2^2d\theta_2^2~.
\label{eq-torusFLAT2}
\end{equation}
By identifying the space under
$\xi\rightarrow \xi+2\pi\Gamma^{-1}$, 
and redefining the coordinates, 
$\Gamma^{-1}\xi=\theta_2$, $\Gamma t=r_1$, we get
\begin{equation}
ds^2=-\Gamma^2dr_1^2+dr_2^2+r_1^2d\theta_1^2+r_2^2d\theta_2^2~.
\label{eq-torusFLAT}
\end{equation}
The conical singularity at $r_1=0$ is not a problem for us;
we do not extend the solution to infinite past, and this
singularity is in the unphysical region, as we will see shortly.


The torus boundary between the de Sitter space and the flat space 
is a $(2+1)$ surface parametrized by $(r_1(\tau),r_2(\tau))$.
The induced metric,
\begin{equation}
ds_{(2+1)}^2=-d\tau^2+r_1^2d\theta_1^2+r_2^2d\theta_2^2~,
\end{equation}
should have the same form, when we approach the domain wall
from either side, (\ref{eq-torusDS}) or (\ref{eq-torusFLAT}).

This condition brings us to the solution
\begin{eqnarray}
r_1(\tau)&=&\frac{1}{\Gamma}\bigg[\varepsilon\sinh(\tau/\varepsilon)
      \pm\sqrt{1+\Gamma^2}\sqrt{l^2-\varepsilon^2}\bigg]~,
       \\ \nonumber
r_2(\tau)&=&\varepsilon\cosh(\tau/\varepsilon)~,
\label{eq-traj}
\end{eqnarray}
as we explain in Appendix~\ref{sec-traj}. 

We take the plus sign in (\ref{eq-traj}), and consider the 
$\tau\ge0$ part to be physical. At $\tau=0$, we have $\dot{r}_2=0$, 
and $r_2$ takes
the minimum value $r_2=\varepsilon$. (We loosely call $\tau=0$ the
nucleation time of the torus bubble.) The parameter $\varepsilon$ 
corresponds to the tension of the domain wall, and $\varepsilon\to 0$ is 
the limit of zero tension domain wall, as we will see below. 

Another parameter $\Gamma$ controls the global time at
the nucleation. The $\Gamma\to 0$ limit corresponds to 
late nucleation, and the $\Gamma\to \infty$ limit corresponds to 
nucleation at the minimal $S^3$, which should correspond to the
case studied in the last subsection. The asymptotic aspect ratio of 
the two circles of the torus is
given by $r_2/r_1=\Gamma$. In the $\Gamma\to \infty$ limit, the ratio
is infinite. $r_1(\tau)$ is constant $r_1(\tau)=\sqrt{l^2-\varepsilon^2}$,
while $r_2(\tau)$ grows to infinite size; this is the situation we
mentioned at the end of the last subsection. 


The junction condition (see Appendix~\ref{sec-junc})
tells us that we need the following
(2+1) dimensional stress tensor on the domain wall:
\begin{eqnarray}
T_\tau^\tau=T_2^2&=&
-2\frac{l-\sqrt{l^2-\varepsilon^2}}{l\varepsilon}
\bigg(1+\frac{\sqrt{1+\Gamma^2}l}{2\Gamma r_1}\bigg)~, \\
T_1^1&=&-2\frac{l-\sqrt{l^2-\varepsilon^2}}{l\varepsilon}~.
\end{eqnarray}
Ordinary domain wall, which is made from a kink of scalar field, 
has $T^{i}_{j}=-\sigma \delta^{i}_{j}$, 
where $\sigma$ is the tension determined by the shape 
of the scalar potential. Here we have extra term proportional
to $1/r_1$, but since it decreases as the torus grows,
we should probably set 
\begin{equation}
\sigma = 2\frac{l-\sqrt{l^2-\varepsilon^2}}{l\varepsilon}~.
\end{equation}

The fact that we need non-standard stress energy on the domain
wall is not surprising. We expect this solution to be an effective
description for a large number of spherical bubbles collided
with each other. In the exact description (as the one in the last
subsection), there is no translational symmetry in one ($r_1$) 
direction. The translational symmetry appears after smearing over the $r_1$ 
direction, but before smearing, there would be defects which
lie along the $r_2$ direction (and are symmetric along $r_2$).
The form of the extra terms, $T^{\tau}_{\tau}=T^{1}_{1}$,
$T^{2}_{2}=0$, is what we expect for such a string-like
object wrapped along $r_2$.

%

\subsection{Boundary with higher topologies}

The above analysis shows the existence of bubbles (true vacuum region) 
with torus boundary. We believe boundaries of any genus can appear
in the radiation shell case as well as in the dust
case discussed at the beginning of this section.

A configuration with genus 2 or larger typically involves ``Y-shape'' 
collisions where three bubbles collide with another bubble in the middle.
To understand the qualitative behavior of this type of collision, we
will use intuition from the analysis of SO(2,1) symmetric 
chain of collisions for the torus case. Local geometry
that results from each collision 
will be roughly the one studied in Section 2.2. Even though we do not have 
the symmetry now, if the geometry is close to flat, we will
be able to use Newtonian approximation, and add the effect of 
each collision. In the middle bubble of the Y-shape collision, 
radiation shells will arrive more often than in the torus case 
(since radiations come from three directions rather than two). 
This may effectively shorten the interval between successive radiation.
(which corresponds to $t_{*(n+1)}-t_{*(n)}$ in Section 2.2).  However, 
this does not change the fact that further radiation shells arrive at an 
increasing interval, which is enough to guarantee that locally the metric
approaches Minkowski space.

The candidate for the asymptotic geometry of the flat region
would be the following: 
\begin{equation}
 ds^2=-dt^2+t^2 ds_{H^3/\Gamma}^2
\end{equation}
where $ds_{H^3/\Gamma}^2$ is the metric of 
a 3-dimensional space obtained by
modding out $H^3$ by elements of a suitable discrete subgroup 
(the Schottky group). See e.g., \cite{krasnov}. 
Negatively curved space with one boundary with 
any genus and arbitrary value of moduli parameters
can be realized this way.
For example, the torus geometry (\ref{eq-torusFLAT2}) 
in the last subsection is equivalently
represented\footnote{Poincar\'{e} coordinates in (\ref{Poincare}) 
are related to the coordinates in (\ref{eq-torusFLAT2}) by 
$x_1=e^{\xi}(r_2/t)\cos\theta_2$,
$x_2=e^{\xi}(r_2/t)\sin\theta_2$,
$z=e^{\xi}\sqrt{t^2-r_2^2}/r_1$. Translation $\xi\to \xi +2\pi\Gamma^{-1}$ 
corresponds to a dilatation by $\lambda=e^{2\pi\Gamma^{-1}}$.} 
as a quotient of $H^3$,
\begin{equation}
 ds^2=-dt^2+t^2 \left(dx_1^2+dx_2^2+ dx_0^2\over x_0^2\right),
\label{Poincare}
\end{equation}
under a scale transformation
$x_i\to\lambda x_i$
with a given $\lambda$ ($|\lambda|>1$).
This transformation has two fixed points, at origin and 
at infinity. 
By further modding out the space by transformations
which have different sets of fixed points and parameters corresponding to 
$\lambda$, we get a higher genus boundary. 
The number of transformations applied corresponds to the 
genus $h$. Any value of the moduli can be realized by choosing
the parameters of the scale transformations~\cite{krasnov}. 
We expect the initial condition produced by bubble collisions
to evolve into this geometry embedded in de Sitter space.
 
The flat region constructed above is causally connected
(i.e. a time like observer in the flat region can eventually 
see the whole region). This is clear from the fact that the whole 
space of an open universe with zero c.c. is causally connected, 
and that taking a quotient only makes causal contact easier. 

This suggests that we should sum over the topology of the 
boundary on which the holographic dual theory is defined. 
We will discuss 
implication of the higher genus boundaries in Section 4. 
Before that, in the next section
we mention an example of geometry which has disconnected
boundaries, which is a little confusing in terms of 
holographic duality.

\section{Multiple boundaries}

Bubble collisions can also produce configurations which have 
multiple boundaries.  
In this case, the dust wall model in Section 2.1 does not guarantee 
the existence of a smooth geometry, since dust walls must cross each other to 
form multiple boundaries. Actually, in 
Section 3.1 we will show that two spherically symmetric 
boundaries must be causally disconnected.  Cases with higher 
genus are less clear and will be discussed in Section 3.2.

\subsection{Two spherical boundaries}

\begin{figure}
\center
\includegraphics[scale=.4]{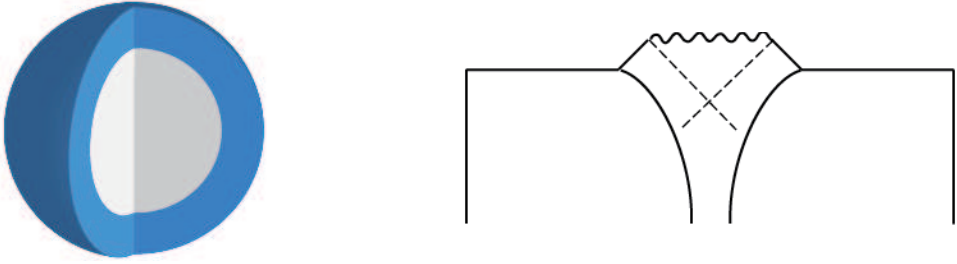}
\caption{Left figure: True vacuum with two spherical boundaries. 
True vacuum is in a (dark blue) shell, and false vacuum fills both sides of
the shell. Right figure: Spacetime which results from this initial 
condition. The two spherical domain walls (assumed to be 
larger than de Sitter horizon) expand monotonically. The flat 
region is described by the Schwarzschild geometry; the 
two domain walls reach different asymptotic flat regions.}
\label{fig-multiple}
\end{figure}

It could happen that bubbles form a ``shell'' rather than a ring
(see Figure~\ref{fig-multiple}). Let us assume that a large
number of bubbles are nucleated on a sphere, and approximate 
the geometry with a spherically symmetric one. 
The space is divided into three regions. The flat region is 
in a thin shell, and it has two disconnected spherical boundaries. 
Let us assume both de Sitter regions are larger than their horizon
size.

This initial condition will evolve into a geometry whose Penrose 
diagram is shown in Figure~\ref{fig-multiple}~\cite{sasaki}. 
From Birkhoff's theorem, in our spherical symmetric situation,
the geometry of the flat region should be the Schwarzschild
geometry. The size of the two spheres (boundaries) should 
increase, since they are larger than de Sitter horizon.
The junction condition tells us that the zero c.c. space is always 
``inside'' de Sitter space, i.e., the former is on the side 
that the area of $S^2$ decreases. Domain wall with such properties
has to be in the 
``white hole'' region of the Schwarzschild geometry.
Patching de Sitter and Schwarzschild geometries, we get
Figure~\ref{fig-multiple}.  

The Schwarzschild mass is determined so that the initial condition
satisfies the junction condition. In the simplest case where
bubbles with negligible size are nucleated along $S^2$ which has
radius $R$ (we assume the $S^2$ is at the ``center'' of $S^3$ of 
global de Sitter), the mass will be~\cite{sasaki}
\begin{equation}
M={R^3\over 2G\ell^2}.
\label{Schwmass}
\end{equation}
We can get this by studying the junction condition 
at the ``nucleation time'' (when the $S^2$ domain walls have
zero radial velocity), in the limit of small tension 
(in this limit the radii of the two $S^2$'s are almost equal to 
$R$)\footnote{The junction condition is given e.g. in~\cite{BGG}. 
In the ``static''
coordinates $(t,r)$ used in~\cite{BGG}, 
the $r$ coordinate (which gives the
size of $S^2$) is timelike in the region of interest 
where $S^2$ is larger than the horizon size. The $S^2$ at rest at the 
center of the global $S^3$ corresponds to $\dot{t}=0$. Substituting this 
into their junction condition, and
setting the domain wall tension zero, we get (\ref{Schwmass}).}.

The two asymptotic regions are separated by horizons. 
A timelike observer who reaches timelike infinity can 
see only one boundary. That observer will feel like there
is a black hole in the interior. 
This black hole will eventually evaporate, leaving
two disconnected geometries. Each geometry has one 
spherical boundary. They will relax to flat space. 

The holographic dual theory is expected to describe
the region that a single observer can see. The above
geometry (at least at late time) will correspond
to a perturbation in the dual theory, similar to the 
one corresponding to 
a black hole in FRW universe.

\subsection{More general cases}
Bubble collisions can also produce configurations which have multiple boundaries
with higher topologies. 

For example, consider two toroidal boundaries:
assume there is a torus inside a torus, and true vacuum fills the
region between the two tori. If this torus is long and thin, 
it can be approximated by infinite straight tubes;
de Sitter space fills inside the inner tube and outside the
outer tube. 
We assume the radii of the inner and outer tubes are both
larger than 
the de Sitter horizon radius. 

It is not obvious how this geometry evolves. The geometry 
of the true vacuum region will not simply be a flat space
with conical deficit, as in the case of cosmic string 
(or the dimensionally reduced 2+1 D gravity). If we
imagine a cosmic string with tension 
given by the energy of the de Sitter region (in the tube 
of radius $R$), its tension would be 
$\mu\sim V_0 R^2 \sim m_p^2H^2R^2>m_p^2$, when the de Sitter region 
is larger than the horizon size $R>H^{-1}$. 
Cosmic string with such a large tension cannot exist, since it
corresponds to a deficit angle larger than $2\pi$. 
In fact, the problem is not purely (2+1) dimensional, since 
we have an extra degree of freedom (the metric component along the tube). 
So the geometry will not be just flat in general. Possibly,
a singularity forms in the true vacuum region, or instability 
occurs and inner de Sitter meet with outer de Sitter.  

A flat region with multiple disconnected boundaries would be confusing in terms of the holographic duality.
This would mean that CFT's defined on each boundary are not 
independent and somehow coupled.
It is not clear how to couple two CFT's without introducing
explicit coupling. This is the issue raised in~\cite{maoz}
for asymptotically Euclidean AdS spaces with multiple boundaries.

We have seen that, at least for the case of spherical boundary,
a timelike observer can see only one boundary. So far we do not
have a clear conclusion for higher genus cases\footnote{We have not
ruled out the possibility that the true vacuum region 
with two genus $h\ge 2$ boundary asymptotes to a  
constant curvature geometry, $H^3$ modded out by ``(quasi-)Fuchsian group,'' 
studied in~\cite{maoz}.}. 
This point clearly deserves further study.

\section{Summing over boundary topologies}

The authors of \cite{FSSY} proposed a holographic dual description
for a bubble with spherical boundary. The proposal is that the dual
theory is a CFT defined on the boundary, and that
the boundary metric (the Liouville field) should be integrated.
Results in the above sections, which show the existence of
boundaries with non-trivial topology,
imply that we have to sum over the topology
of the base space on which the dual CFT is defined.
This suggests that eternal inflation is described by
a kind of ``string theory.''  The CFT has $c \gg 26$ and is 
coupled to Liouville, and so it is a ``supercritical'' string theory.

How should each topology be weighted? i.e., what is the
string coupling $g_s$?
Adding a handle requires a minimal number of extra bubbles,
$k$ (which might  be two or three). It seems
appropriate to identify the coupling constant as
$g_s\sim \gamma^k$.

From the string theory point of view there are a number of peculiarities
in the sum over topologies.  First, to nucleate a torus with a modulus
$\tau$ corresponding
to a large aspect ratio ($\tau_2 \sim r_2/r_1$) requires many bubbles to be
nucleated.  This means that this region
of moduli space is strongly suppressed by a factor that looks roughly like
$g_s^{\tau_2}$. This is surprising. 
``Pseudotachyons'' typically appear in supercritical
strings~\cite{silverstein, hellerman}.
They will cause an IR divergence at $\tau_2\to \infty$. Also
there is no extra $g_s$ dependence in the genus one amplitude.

From the bubble nucleation point of view it seems that the sum over higher
topologies is convergent.
All configurations of $n$ connected bubbles can be thought of as a branched
polymer with $n$ nodes.
There are order $\exp(\alpha n)$ such configurations (where
$\alpha$ is a constant of order one).   
So the sum over them
appears to be convergent.
The higher genus contributions are a systematically small part of the sum.

This is surprising from the string theory point of view.  There the integral
over moduli space of genus $h$
typically goes like $(2h)!$ indicating a divergent series and
characteristic nonperturbative effects of size  $e^{-C/g_s}$~\cite{shenker}.

We do not have a full understanding of these differences.  We can point to
one novel aspect of the
``string theory'' of~\cite{FSSY} 
which might be relevant.  
The central charge is argued to be $c \sim S$
where $S$ is the ancestor de Sitter entropy~\cite{FSSY, censustaker}. 
Nucleation rate is $\gamma \geq \exp(-S)$.  So,
roughly speaking, $g_s \sim \exp(-c)$.  The world-sheet parameters of 
the string theory are linked to the string coupling.

\section*{Acknowledgements}
We thank Tom Banks, Simeon Hellerman and Eva Silverstein for helpful
discussions.
The work of YS is supported in part by MEXT 
Grant-in-Aid for Young Scientists (B) No.19740173.
SS, LS, and CY are supported in part by NSF Grant 0244728.
RB, BF, and IY are supported by the
Berkeley Center for Theoretical Physics, by a CAREER grant of the 
National Science Foundation, 
and by DOE grant DE-AC0376SF00098. 

\appendix

\section{Domain wall trajectory}
\label{sec-traj}
In this appendix, we solve the Israel junction conditions to find a solution which consists of a toroidal domain wall separating de Sitter space from Minkowski space.

The trajectory of the domain wall $r_i(\tau)$ satisfies the 
following two equations:
\begin{eqnarray}
-1&=&-\Gamma^2\dot{r}_1^2+\dot{r}_2^2~, 
\label{eq-flat} \\
-1&=&\frac{(r_2^2-l^2)\dot{r}_1^2+(r_1^2-l^2)\dot{r}_2^2
      -2r_1r_2\dot{r}_1\dot{r}_2}{r_1^2+r_2^2-l^2}~,
\end{eqnarray}
where the dot denotes the derivative w.r.t. the proper time $\tau$.

We can combine them and use $dr_2/dr_1=\dot{r}_2/\dot{r}_1$ to 
get a first order differential equation 
\begin{equation}
r_2^2\bigg(\frac{dr_2}{dr_1}\bigg)^2+2r_1r_2\frac{dr_2}{dr_1}
-[\Gamma^2r_1^2+(1+\Gamma^2)(r_2^2-l^2)]=0~.
\end{equation}
Solving it as a quadratic equation first, we have
\begin{equation}
r_2\frac{dr_2}{dr_1}
+r_1=\pm\sqrt{(1+\Gamma^2)(r_1^2+r_2^2-l^2)}~.
\end{equation}
Changing the variable to $Q=r_1^2+r_2^2$, 
we get a simple differential equation,
\begin{equation}
\frac{dQ}{dr_1}=\pm2\sqrt{1+\Gamma^2}\sqrt{Q-l^2}~, 
\end{equation}
with the solution
\begin{equation}
\sqrt{Q-l^2}=\pm\sqrt{1+\Gamma^2}(r_1+c)~,
\end{equation}
where $c$ is an integral constant. We will take the plus sign,
since this corresponds to the case of interest
where both $r_1$ and $r_2$ are growing.

This provides the trajectory equation for $r_1$ and $r_2$,
\begin{equation}
\bigg[\Gamma r_1+\frac{1+\Gamma^2}{\Gamma}c\bigg]^2-r_2^2
=-(l^2-c^2\frac{1+\Gamma^2}{\Gamma^2})~.
\end{equation}
Together with Eq.~(\ref{eq-flat}), we can see that 
\begin{eqnarray}
\Gamma r_1+\frac{1+\Gamma^2}{\Gamma}c &=& 
\sqrt{l^2-c^2\frac{1+\Gamma^2}{\Gamma^2}}
\sinh\frac{\tau}{\sqrt{l^2-c^2\frac{1+\Gamma^2}{\Gamma^2}}}~, 
\nonumber \\
r_2 &=& \sqrt{l^2-c^2\frac{1+\Gamma^2}{\Gamma^2}}
\cosh\frac{\tau}{\sqrt{l^2-c^2\frac{1+\Gamma^2}{\Gamma^2}}}~.
\end{eqnarray}
It is convenient to define
\begin{equation}
\varepsilon=\sqrt{l^2-c^2\frac{1+\Gamma^2}{\Gamma^2}}~,
\end{equation}
which gives us Eq.~(\ref{eq-traj}). 

\section{Junction conditions}
\label{sec-junc}
In this appendix we review the Israel junction conditions and compute the extrinsic curvature for the case of interest, a toroidal domain wall.

The $(2+1)$D stress tensor of the domain wall is related to the jump
in extrinsic curvature (see e.g.,~\cite{BGG}), 
\begin{equation}
T_{\mu}^{\nu}={\rm Tr}(\Delta K)\delta_{\mu}^{\nu}-\Delta K_{\mu}^{\nu}~,
\end{equation}
\begin{equation}
\Delta K_{\mu\nu}=K_{\mu\nu}^{\rm dS}-K_{\mu\nu}^{\rm flat}~.
\end{equation}

To calculate the extrinsic curvature, it is convenient to write down 
the Gaussian normal coordinate in the vicinity of the domain wall.
Suppressing the symmetric directions $(\theta_1,\theta_2)$, we need 
a coordinate transformation ${\rm(r_1,r_2)}\rightarrow(\tau,\eta)$.
\footnote{Here (and only here) we use the printed ${\rm r_i}$ to 
denote $r_i$ as the coordinates (independent variables), as opposed to
$r_i(\tau)$, which is a given function of the coordinate $\tau$. 
In the following, we will abbreviate the latter as $r_i$,
since we believe its meaning is clear from the context.} 
\begin{equation}
{\rm r_i}=r_i(\tau)+v_i(\tau)\eta+O(\eta^2),
\label{eq-trans}
\end{equation}
where $\eta=0$ is the domain wall and $(v_1,v_2)$ is the unit 
vector orthogonal to it.  

On the flat space side, we have
\begin{eqnarray}
v_1(\tau)&=&\dot{r}_2(\tau)/\Gamma~, \nonumber \\
v_2(\tau)&=&\dot{r}_1(\tau)\Gamma~,
\end{eqnarray}
and the Gaussian normal coordinate is 
\begin{equation}
ds^2=d\eta^2-\{\Gamma^2[\dot{r}_1+\dot{v}_1\eta]^2-
[\dot{r}_2+\dot{v}_2\eta]^2\}d\tau^2+
[r_i+v_i\eta]^2d\theta_i^2~.
\label{eq-gauflat}
\end{equation}
We ignore the higher order $\eta$ terms in Eq.~(\ref{eq-trans}), 
which will not be needed for the calculation of the extrinsic
curvature,
\begin{equation}
K_{ij}=\frac{1}{2}
\frac{\partial g_{ij}}{\partial\eta}\bigg|_{\eta=0}~.
\label{eq-ext}
\end{equation}
From (\ref{eq-gauflat}) and (\ref{eq-ext}), we get
\begin{eqnarray}
K_{\tau\tau}^{\rm flat}&=&
           -\Gamma^2\dot{r}_1\dot{v}_1
           +\dot{r}_2\dot{v}_2~, \nonumber \\
K_{11}^{\rm flat}      &=&    r_1v_1~, \nonumber \\
K_{22}^{\rm flat}      &=&    r_2v_2~.
\end{eqnarray}

On the de Sitter side, the normal vector takes a different form,
\begin{eqnarray}
u_1(\tau) &=& \frac{(r_1^2-l^2)\dot{r}_2-r_1r_2\dot{r}_1}
                   {l\sqrt{r_1^2+r_2^2-l^2}}~, \nonumber \\
u_2(\tau) &=&-\frac{(r_2^2-l^2)\dot{r}_1-r_1r_2\dot{r}_2}
                   {l\sqrt{r_1^2+r_2^2-l^2}}~.
\end{eqnarray}
From (\ref{eq-trans}) and (\ref{eq-gauflat}) with $v_i$
replaced by $u_i$, we get
\begin{eqnarray}
K_{11}^{\rm dS}      &=&    r_1u_1~, \nonumber \\
K_{22}^{\rm dS}      &=&    r_2u_2~,
\end{eqnarray}
and $K_{\tau\tau}^{\rm dS}$ as a formidable combination of 
$(r_i,\dot{r}_i,u_i,\dot{u}_i)$:
\begin{eqnarray}
K_{\tau\tau}^{\rm dS} &=&
         \frac{\partial q_{11}}{2\partial r_1}\dot{r}_1^2u_1
        +\frac{\partial q_{11}}{2\partial r_2}\dot{r}_1^2u_2
        +q_{11}\dot{r}_1\dot{u}_1~, \nonumber \\
      &+&\frac{\partial q_{22}}{2\partial r_1}\dot{r}_2^2u_1
        +\frac{\partial q_{22}}{2\partial r_2}\dot{r}_2^2u_2
        +q_{22}\dot{r}_2\dot{u}_2~, \nonumber \\
      &+&\frac{\partial q_{12}}{\partial r_1}\dot{r}_1\dot{r}_2u_1
        +\frac{\partial q_{12}}{\partial r_2}\dot{r}_1\dot{r}_2u_2
        +q_{12}(\dot{u}_1\dot{r}_2+\dot{r}_1\dot{u}_2)~.
\end{eqnarray}
Here $q_{ij}$ is the metric component in Eq.~(\ref{eq-torusDS}),
\begin{eqnarray}
q_{11}&=&\frac{r_2^2-l^2}{r_1^2+r_2^2-l^2}~, \nonumber \\
q_{22}&=&\frac{r_1^2-l^2}{r_1^2+r_2^2-l^2}~, \nonumber \\
q_{12}&=&\frac{-r_1r_2}{r_1^2+r_2^2-l^2}~.
\end{eqnarray}

Combining Eq.~(\ref{eq-traj}) with the extrinsic curvature in this 
section, we have
\begin{eqnarray}
\Delta K_{\tau}^{\tau}=\Delta K_2^2 &=& 
-\frac{l-\sqrt{l^2-\varepsilon^2}}{\varepsilon l}~, \nonumber \\
\Delta K_1^1 &=& - \frac{l-\sqrt{l^2-\varepsilon^2}}{\varepsilon l}
\bigg(1\pm\frac{l\sqrt{1+\Gamma^2}}{r_1\Gamma}\bigg)~.
\end{eqnarray}
The $\pm$ signs correspond to those in (\ref{eq-traj}).

\end{document}